\documentclass[]{spie}  
\usepackage[]{graphicx}

\title{Upgrading the processing pipeline for the National Park Service Night Skies Program} 

\author{Li-Wei Hung\supit{a}, Davyd Betchkal\supit{a}, Sharolyn J. Anderson\supit{a}, and Damon Joyce\supit{a}
\skiplinehalf
\supit{a}National Park Service, Natural Sounds and Night Skies Division, 1201 Oakridge Drive, Suite 100, Fort Collins, Colorado, USA
}

\authorinfo{Further author information: (Send correspondence to Li-Wei Hung)\\Li-Wei Hung: E-mail: li-wei\_hung@nps.gov, Telephone: 1 970 267 2196}

\begin{document} 
  \maketitle 

\begin{abstract}

The US National Park Service (NPS) assesses the night sky quality over parks by capturing a series of overlapping images to obtain a mosaic view of the entire night sky. The NPS Night Skies Program has integrated a sequence of scripts and software packages --- a ``pipeline" --- to process and create the hemispherical mosaic images. This processing pipeline is being improved to reduce dependence on commercial software packages, improve management of revisions, and ease distribution of the latest version. The upgraded pipeline is designed for processing the images in three stages: (I) performing data reduction, calibration, and mosaic, (II) modeling the natural sky brightness to separate out light from artificial sources, and (III) deriving sky quality indicators. Currently, stage I is completed and is presented in detail in this report. Stage II and III are in the upgrading process and will be presented in a future report. In stage I, the pipeline applies basic image reduction, pointing registration, photometric calibration, coordinate transformation, and image mosaicking. We implemented new features including auto-logging the processing history, version control through Git and GitHub, team management on source code development, reduction on the number of required proprietary software usage, and setting the primary pipeline language to Python. Once the upgrade is completed, our open source pipeline can benefit other scientists in the similar research field worldwide for processing related sets of data.
\end{abstract}


\section{INTRODUCTION}
\label{sec:intro}  

The US National Park Service (NPS) perceives the natural night sky as an inseparable element of wild areas for park visitors and wildlife. The night sky is a natural, cultural, educational, and economic resource. The Natural Sounds and Night Skies Division, in collaboration with NPS regions, parks, and programs, provides service-wide support for night sky and nocturnal resource conservation through measurements, modeling, critical analysis, knowledge synthesis, and informed decision making.
Our first step towards understanding and protecting the night sky is to assess the night sky quality. The NPS developed the camera system composed of a Nikon lens, a V-band filter, and a CCD camera \cite{duri07}. The system captures a series of overlapping images to obtain a high-resolution mosaic of the entire sky from horizon to horizon. These images allow us to assess the condition of the night sky and the size and intensity of artificial light domes.

The NPS Night Skies Program has integrated a sequence of image processing software --- the ``pipeline" --- to merge and process the 45 one-million-pixel images taken by our mobile camera system. Specifically, the pipeline applies basic reduction, performs photometric calibration, mosaics the images to form the panoramic views of the entire night sky, generates the models for the natural sky brightness \cite{duri13}, and calculates values of various sky brightness metrics \cite{duri16}. This original data pipeline was developed mainly by Dan Duriscoe about a decade ago. These scripts were written in three different languages including Python, Java, and Visual Basic. They interact with six different proprietary software packages including ACP Observatory Control, ArcGIS, MaxIm DL, Microsoft Excel, Photoshop, and PixInsight. Although the current image processing pipeline is functional, an upgrade is underway to make the pipeline distributable and manageable. The upgraded pipeline is designed for processing the images in three stages: (I) performing data reduction and calibration, (II) modeling the natural sky brightness to separate out light from artificial sources, and (III) deriving sky quality indicators. Stage I is completed. Stage II and III are in the upgrading process and will be presented in a future report.

This report focuses on stage I of the pipeline. We describe the detailed data reduction and calibration processes in Sec.~\ref{sec:stage1}. Next, we highlight new features implemented on this upgraded pipeline in Sec.~\ref{sec:features}. We then summarize the project in Sec.~\ref{sec:summary}. Once the entire upgrade is completed, the pipeline will be available through GitHub to NPS staff, research partners, and other scientists who may adapt the software to achieve other objectives. 

\section{DATA REDUCTION AND CALIBRATION}
\label{sec:stage1} 

\begin{figure}
   \begin{center}
   \begin{tabular}{c}
   \includegraphics[height=9cm]{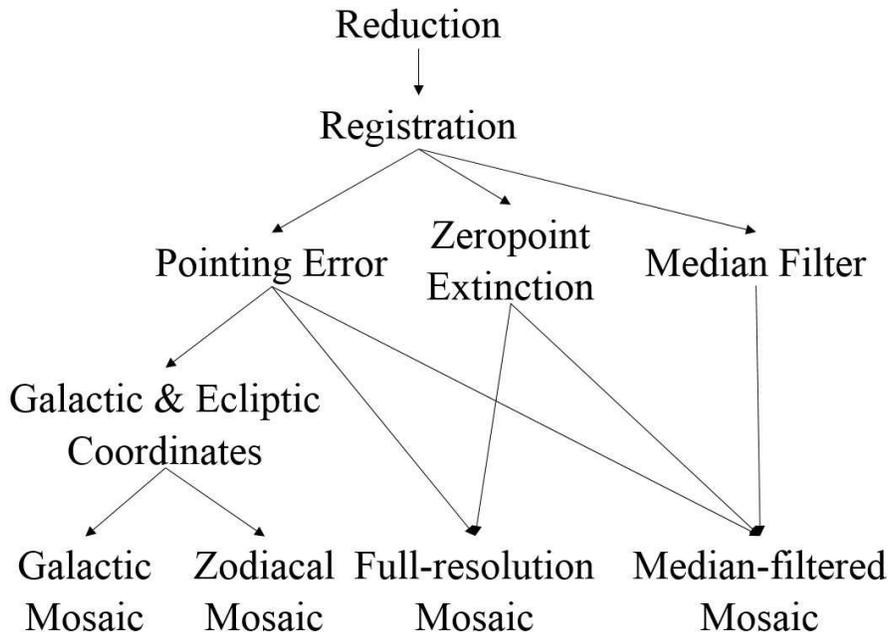}
   \end{tabular}
   \end{center}
   \caption[flowchart] 
   { \label{fig:flowchart} 
Data reduction and calibration flowchart. During this stage, the pipeline takes the raw images as the inputs, applies basic noise reduction, determines the image pointing, performs absolute photometry, and mosaics images to form panoramic views. Procedures are executed in-parallel if they are independent from each other.}
   \end{figure}

The first processing stage of the pipeline focuses on data reduction and calibration. Specifically, it applies basic image reduction, pointing registration, photometric calibration, coordinate transformation, and image mosaicking. Fig.~\ref{fig:flowchart} shows the relationship between these procedures. We describe the purpose and the method used for each procedure in detail in the following subsections. The pipeline uses all the available processors in a single computer. Each procedure begins as soon as all the dependent procedures have finished. Independent procedures are executed in-parallel to maximize the computing efficiency. 

\subsection{Reduction}
\begin{figure}
   \begin{center}
   \begin{tabular}{c}
   \includegraphics[width=\linewidth]{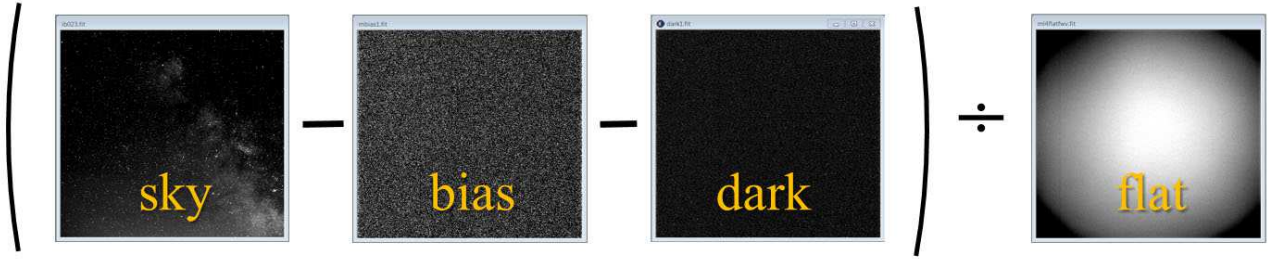}
   \end{tabular}
   \end{center}
   \caption[reduction] 
   { \label{fig:data_reduction} 
Simplified illustration of the data reduction implemented in the pipeline. Each sky (science) image is bias and dark subtracted and then divided by the flat. Correction for the non-linearity response of the detector is not shown here but is applied at various places as described in the text.}
   \end{figure}

This procedure takes raw images as input and performs basic image reduction. This reduction includes corrections for bias, dark, flat, and non-linearity response of the detector. Fig.~\ref{fig:data_reduction} shows a simplified illustration of the data reduction implemented in the pipeline. Correction for the non-linearity response of the detector is not shown here but is applied at various places as described later in the text.

Flat-fielding $F$ and detector response curve $L$ are characterized in lab. Flats are used to correct for different gains to ensure the uniformity of pixel-to-pixel sensitivity. The detector response curve is used for correcting the non-linearity response of detector, especially at the low signal region and near saturation. At the beginning of data collection for each data set, five dark $D$ images and five bias $B$ images are taken in the field. Dark and bias images are used for characterizing the thermal noise and the read noise from the CCD detector correspondingly. 

Each dark image is then processed as following to obtain the calibrated dark $D_c$ image: $D_c = (D - B) * L$. Next, the script creates the master dark image $D_m$ through averaging the five $D_c$. Similarly, the script creates the master bias image $B_m$ through averaging the five biases. While observing, an additional 50 pixels by 50 pixels small bias image $B_s$ is taken immediately after each sky image $S$. We use these small bias images to track the bias drift $B_d$ over the course of the observation. We compute the bias drift by subtracting the average value of the central 50 pixels by 50 pixels of the master bias from the average pixel value of each $B_s$. Finally, to obtain a calibrated sky image $S_c$, we use the following equation: 
	\begin{equation}
	\label{eq:reduction}
S_c = \frac{(S - B_m - B_d) \times L - D_m}{F} .
	\end{equation}
All of the terms in the above equation are 2D image arrays except for $B_d$ which is a single-valued scale factor. 
This script outputs calibrated science images in both .fits and .tif formats.

\subsection{Registration} 
\begin{figure}
   \begin{center}
   \begin{tabular}{c}
   \includegraphics[width=0.7\linewidth]{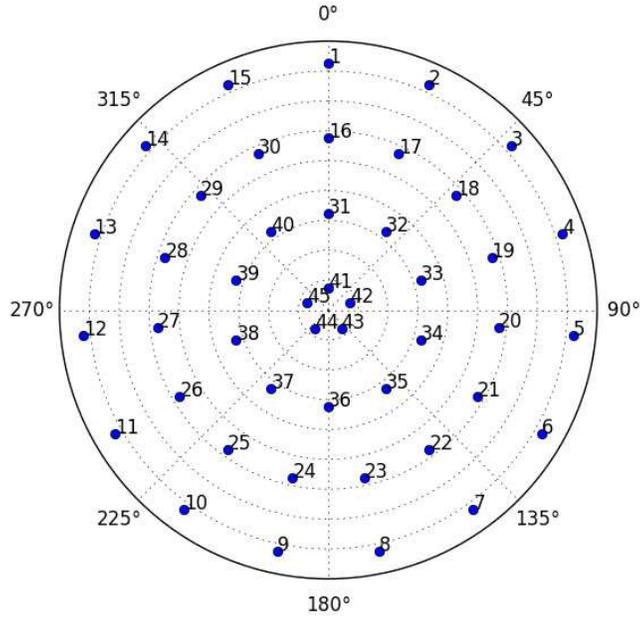}
   \end{tabular}
   \end{center}
   \caption[reduction] 
   { \label{fig:pointing} 
The commanded camera pointing positions plotted in the azimuth-altitude coordinate system. Each image covers 24$^\circ$ by 24$^\circ$ of the sky. A set of 45 images are taken to cover the entire sky.}
   \end{figure}

During the data collection, a set of images are taken to cover the entire sky. Fig.~\ref{fig:pointing} shows the commanded camera pointing positions plotted in the azimuth-altitude coordinate system for one of our camera systems. Although the camera movement is controlled by a computer, the actual camera pointing often deviates slightly from the exact commended position. This processing procedure registers the pointing of the images to the sky coordinates by matching the position of standard stars captured in the image to the position of the stars listed in the Tycho-2 catalog. Once the pointing is determined, the image will register the updated coordinates in the header.
   
\subsection{Pointing Error}
As the name suggested, this procedure calculates the pointing error by comparing the commended pointing position to the actual pointing position found in the previous step. The sky coordinates and the pointing error are interpolated for the images that failed to be registered using the standard stars. The correction computed from this script is later used to accurately mosaic the images together. 

\subsection{Zeropoint \& Extinction}
\begin{figure}
   \begin{center}
   \begin{tabular}{c}
   \includegraphics[width=0.7\linewidth]{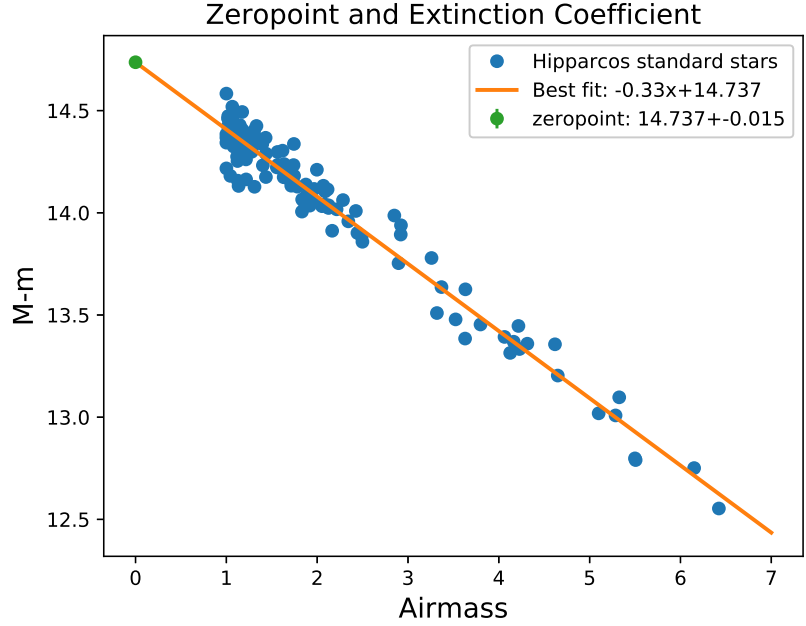}
   \end{tabular}
   \end{center}
   \caption[reduction] 
   { \label{fig:zeropoint} 
Zeropoint and extinction values derived from standard stars (blue dotes) on an example data set. We first plot the difference between the absolute magnitude $M$ and the apparent magnitude $m$ against the airmass for all the standard stars passed through the selection threshold. Then, we fit a straight line to obtain the instrumental zeropoint (y-intersect) and the extinction coefficient (slope).}
   \end{figure}

This procedure derives the best-fit extinction coefficient and the instrumental zeropoint needed for performing absolute photometric calibration. For each standard star identified in the image, we first measure the background brightness through aperture photometry using a ring-shaped aperture with the radius extending from 4 to 8 pixels. Then we subtract the averaged background value and fit the background-subtracted pixels located within 3 pixels of the star center with a 2D Gaussian. We only record and keep the star if it passes all these thresholds: (a) the $\sigma$ of the best-fit Gaussian must be less than 2 pixels, (b) the signal to noise ratio must be greater than 25, and (c) the best-fit stellar centroid must be within 1 pixel of the position listed in the catalog. Next, we convert the measured flux $F$ in [DN/s] to the apparent magnitude $m$ using this equation: $m = -2.5 \times log(F)$. Separately, this script computes the airmass of the stars according to their altitude. Once the brightness and airmass is measured for each standard star, we plot the difference between the absolute magnitude $M$ and the apparent magnitude $m$ against the airmass. Finally, we fit the points with a straight line using the least-square method through numpy.polyfit in Python. From the fit results, we obtain the instrumental zeropoint (y-intersect), the extinction coefficient (slope), and their uncertainties. Fig.~\ref{fig:zeropoint} shows the output graph from an example data set. 

\subsection{Medium Filter}
Because the main interest for collecting these images is to measure the sky background brightness, we apply a median filter to each image to remove point sources and stars in this procedure. The filtered images will only show the diffused light. The optimal filter size depends on the plate scale of the image and the size of the point spread function. The filter size should be selected to ensure most point sources are effectively filtered out, and the median value in that filter window is a good representation of the background brightness. In the script, we set the default filter mask radius to be 0.5 degree. 

\subsection{Galactic \& Ecliptic Coordinates}
This procedure calculates ecliptic coordinate, galactic coordinate, and rotation angles of each image. The script reads in the registered image coordinates in azimuth and altitude along with the longitude and latitude of the observing site from the image header. Then, it uses the ACP Observatory software to calculate the corresponding galactic and ecliptic coordinates and the rotation angles of the images relative to the galactic and zodiacal planes. These output coordinates will be used in producing the zodiacal light mosaic and the Milky Way mosaic in the later process specifically for the given location, date, and time of the data set.

\subsection{Galactic Mosaic}
\begin{figure}
   \begin{center}
   \begin{tabular}{c}
   \includegraphics[width=\linewidth]{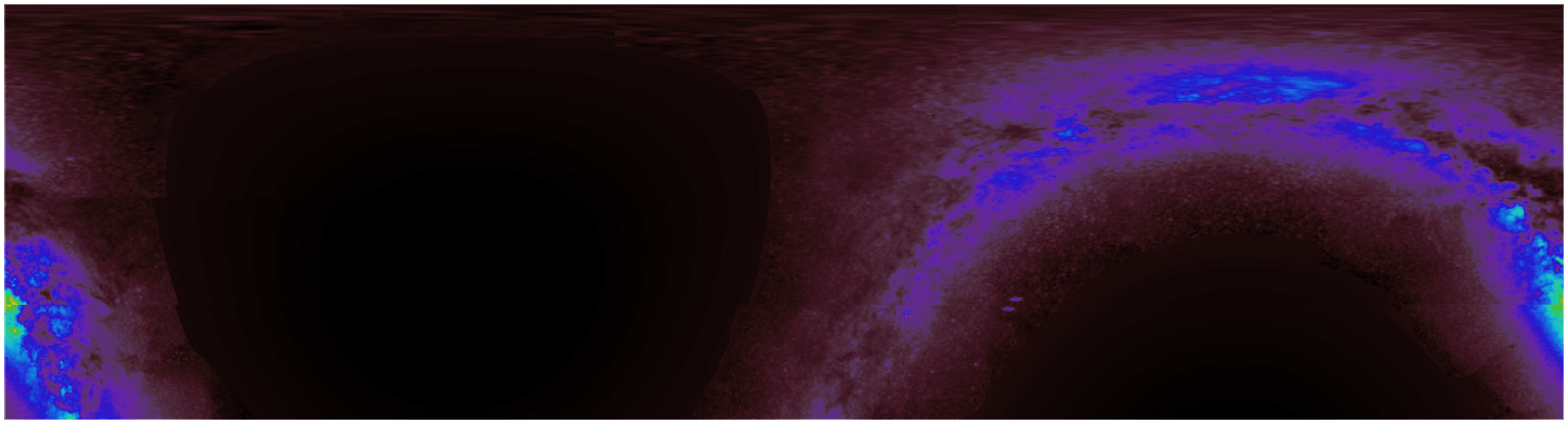}\\
   \includegraphics[width=\linewidth]{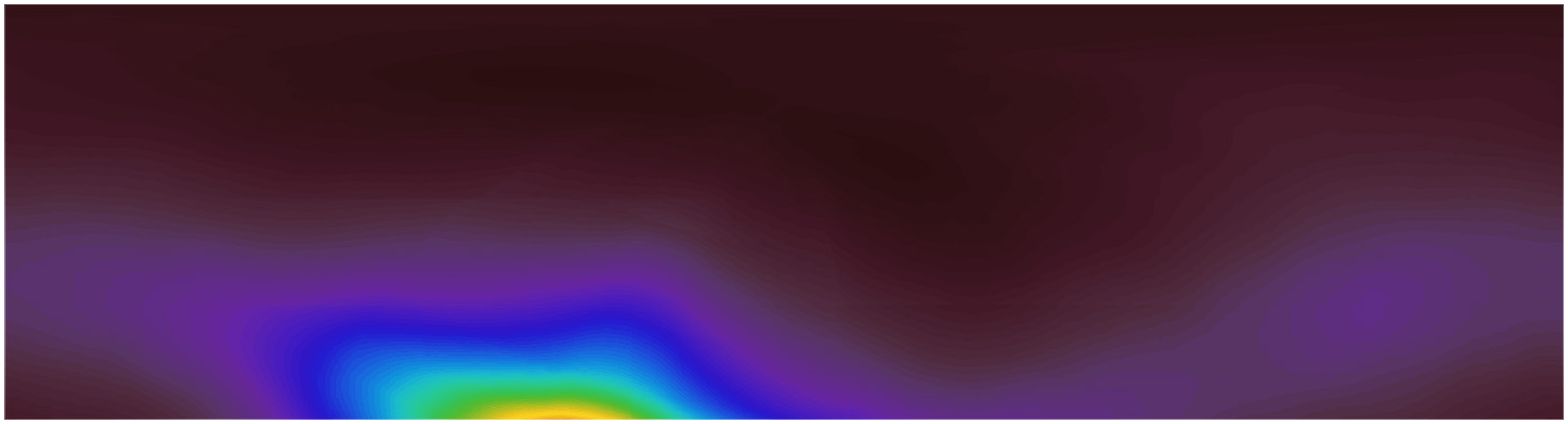}\\
   \includegraphics[width=\linewidth]{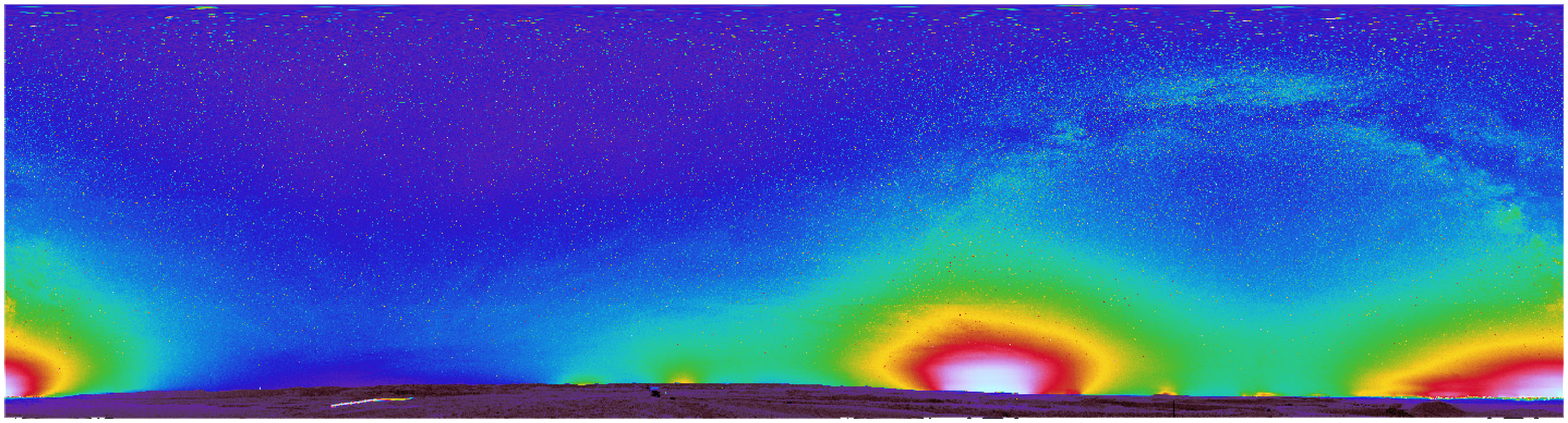}\\
   \includegraphics[width=\linewidth]{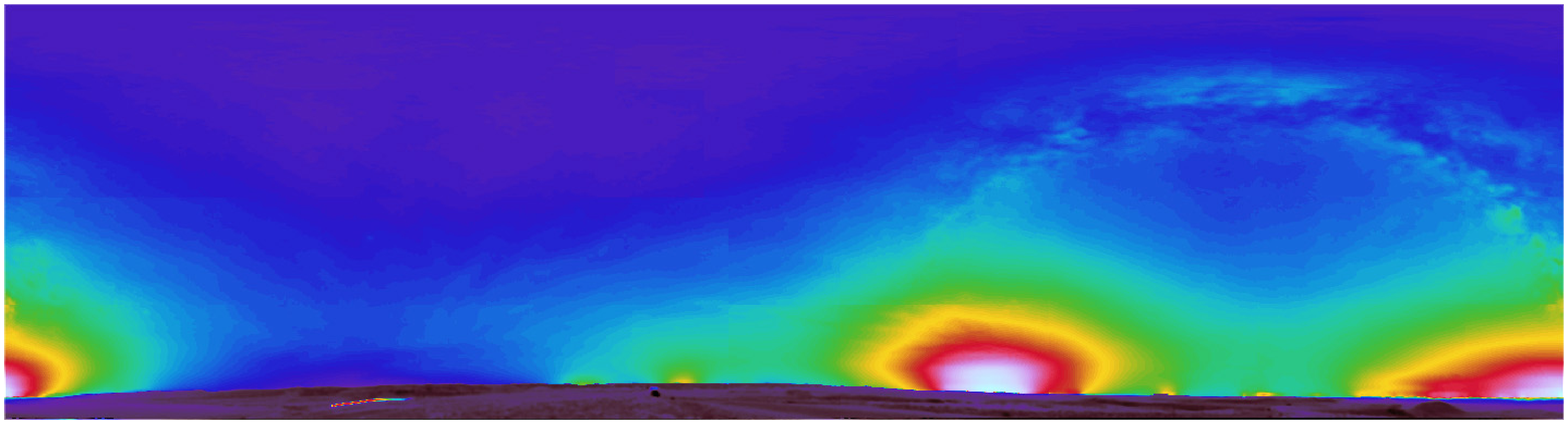}
   \end{tabular}
   \end{center}
   \caption[] 
   { \label{fig:mosaics} 
The panoramic mosaics of an example data set. The horizontal axis is azimuth and the vertical axis is altitude. All of these panoramas are photometrically calibrated. From the top: the Milky Way model, the zodiacal light model, full-resolution mosaic of the observed night sky, and median-filtered mosaic of the observed night sky. Every Milky Way model and zodiacal light model is constructed specific to the time and the location of the observed data set. Several light domes are visible along the horizon in this example data set.}
   \end{figure}

In this procedure, a Milky Way model is constructed based on the time and the location of the observed data set. We use arcpy to generate this model in ArcGIS. This script first reads in a raster file of the Milky Way model template\cite{duri13} in the galactic plane. The script then uses the previously determined galactic coordinates and the rotation angles to model the appearance of the Milky Way in each image. We then correct the distortion from the projection and piece them to form the panoramic view of the Milky Way model specific to the observed data set. The top panel in Fig.~\ref{fig:mosaics} shows the Milky Way model constructed for an example data set. The horizontal axis is azimuth and the vertical axis is altitude.  

\subsection{Zodiacal Mosaic}
In this procedure, a zodiacal light model is constructed based on the time and the location of the observed data set. We use arcpy to generate this model in ArcGIS. This script first reads in a raster file of the zodiacal light model template\cite{duri13} in the ecliptic plane. The script then uses the previously determined ecliptic coordinates and the rotation angles to model the appearance of the zodiacal light in each image. We then correct the distortion from the projection and piece them to form the panoramic view of the zodiacal light model specific to the observed data set. The second panel in  Fig.~\ref{fig:mosaics} shows the zodiacal light model constructed for an example data set.

\subsection{Full-resolution Mosaic}
In this procedure, a full-resolution panoramic view of the sky is constructed form the observed images. We use arcpy to project, remove distortion, clip, and mosaic images. This script first reads in the full resolution tiff files, corrects for the pointing error, mosaics them to form the panoramic image, and then perform absolute photometric calibration to convert the mosaic from the raw unit in data number (DN) to calibrated unit in magnitudes per square arcsecond using the following equation:
	\begin{equation}
	\label{eq:photometry}
M_c = Z - 2.5 \times log(M_r / t / P^2),
	\end{equation}
where $M_c$ is the mosaic in calibrated unit of magnitudes per square arc second, $Z$ is the instrumental zeropoint in magnitude, $M_r$  is the mosaic in raw unit of DN, $t$ is the exposure time in seconds, and $P$ is the plate scale in arcsecond per pixel. The third panel in Fig.~\ref{fig:mosaics} shows the final panoramic image in full resolution of an example data set output from this process.

\subsection{Median-filtered Mosaic} 
This script makes the whole sky mosaic from the median filtered images. This procedure is the same as the procedure for making the full-resolution mosaic except the input images are the median-filtered images. The bottom panel in Fig.~\ref{fig:mosaics} shows the panoramic image constructed from the median-filtered images. Note that stars and point sources are removed in this image, leaving only the calibrated background sky brightness in magnitudes per square arc second.

\section{FEATURES OF THE NEW PIPELINE}
\label{sec:features} 
\subsection{Software Requirement} 
The original pipeline scripts are written in three different languages including Python, Java, and Visual Basic and interact with six different proprietary software packages including ACP Observatory Control, ArcGIS, MaxIm DL, Microsoft Excel, Photoshop, and PixInsight. Currently, we are able to upgrade the pipeline to perform the same procedures without using Java, Visual Basic, Microsoft Excel, and PixInsight. Python is our main scripting language for the upgraded pipeline. 

\subsection{Processing History Log}
\begin{figure}
   \begin{center}
   \begin{tabular}{c}
   \includegraphics[width=\linewidth]{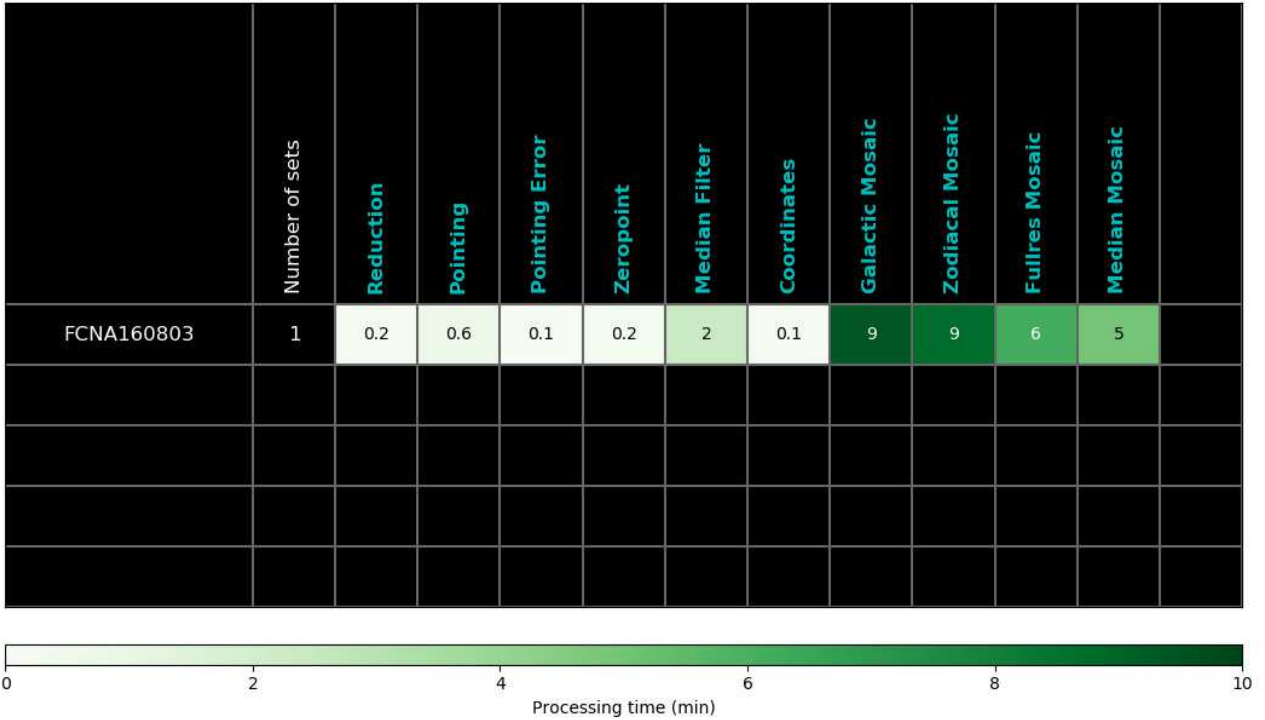}
   \end{tabular}
   \end{center}
   \caption[] 
   { \label{fig:time} 
Processing time spent for each procedure on an example data set. Mosaic procedures usually take the longest to finish. After the pipeline finish running, the graph will be saved as part of the processing record.}
   \end{figure}

The upgraded pipeline logs the user input on various parameters required for running the pipeline. Some of the example parameters include the name and parts of the data to be processed and the specific calibration files used. The log also records the processing history from the intermediate outputs of different procedures. Finally, it records the total processing time as well. In additional to the text log, a real-time graph will pop out to show the progress of processing. After the pipeline finish running, the graph will be saved as part of the processing record. Fig.~\ref{fig:time} shows an example of the processing time graph saved after the pipeline finishes running. 

\subsection{Source Code Management}
We use Git and GitHub for version control, project management, and source-code distribution. The original data pipeline was developed mainly by a single person about a decade ago. Introducing Git and GitHub allows us to transition from a single-person initiated top-down model to a bottom-up model for implementing any updates or new development. Our project is open source and is avalible for download on the GitHub page: https://github.com/liweihung/nightskies.

\section{SUMMARY}
\label{sec:summary} 

The US National Park Service (NPS) assesses the night sky quality in the parks by capturing a series of overlapping images to obtain a mosaic view of the entire night sky. The NPS Night Skies Program has specialized pipeline to perform data reduction and processing of these 45 one-million-pixel images taken by our mobile camera system. Although the current image processing pipeline is functional, an upgrade is underway to make the pipeline distributable and manageable. The upgraded pipeline is designed for processing the images in three stages: (I) performing data reduction and calibration, (II) modeling the natural sky brightness to separate out light from artificial sources, and (III) deriving sky quality indicators. Specifically, in stage I, the pipeline applies basic image reduction, pointing registration, photometric calibration, coordinate transformation, and image mosaicking. Currently, stage I is completed, stage II and III are in the upgrading process and will be presented in a future paper. 

The upgraded pipeline is completely written in Python and reduces the number of required proprietary software packages. Some new pipeline features include auto-logging user-input parameters, the processing history, and the processing time. We are using Git and GitHub for version control, project management, and source-code distribution. Introducing Git and GitHub to the upgraded pipeline allows us to transition from a single-person initiated top-down model to a bottom-up model for implementing any future pipeline updates or new development. Once the entire upgrade is completed, the pipeline will be managed and distributed via GitHub to other NPS staff, partners, and other scientists. All of our Python pipeline scripts will be open source where we hope to also benefit other scientists in the similar research field worldwide.


\acknowledgments  
The authors acknowledge Dan M. Duriscoe for clarifying questions regarding the original pipeline and providing feedback on the upgraded pipeline. We also want to thank Kurt Fristrup for reviewing this report and providing constructive comments.  


\newcommand{\noopsort}[1]{}

\bibliographystyle{spiebib}   

\end{document}